# Novel Multiferroic Nanocomposite with High Pressure-Modulated Magnetoelectric Coupling

Chunrui Song[a], Nikita Liedienov[a,b], Igor Fesych[c], Roman Kulagin[d], Yan Beygelzimer[b], Xin Zhang[a], Yonghao Han[a,*], Quanjun Li[a], Bingbing Liu[a], Aleksey Pashchenko[a,b,e,*], Georgiy Levchenko[a,b,*]

[a] *State Key Laboratory of Superhard Materials, International Center of Future Science, Jilin University, Changchun 130012, P. R. China*
[b] *Donetsk Institute for Physics and Engineering named after O.O. Galkin, NAS of Ukraine, Kyiv 03028, Ukraine*
[c] *Taras Shevchenko National University of Kyiv, Kyiv 01030, Ukraine*
[d] *Institute of Nanotechnology, Karlsruhe Institute of Technology, Karlsruhe 76021, Germany*
[e] *Institute of Magnetism, NAS of Ukraine and MES of Ukraine, Kyiv 03142, Ukraine*

These authors contributed equally: Chunrui Song, Nikita Liedienov

*Corresponding author
*E-mail address*:   hanyh@jlu.edu.cn (Yonghao Han)
          alpash@ukr.net   (Aleksey Pashchenko)
          g-levch@ukr.net (Georgiy Levchenko)

## Abstract

In this work, we have designed obtained a novel multiferroic nanocomposite using the high-pressure torsion (HPT) method. The crystal structure, phase composition, morphology, ferromagnetic (FM) and ferroelectric (FE) properties of the initial powders and "multiferroic-ferromagnetic" nanocomposites have been studied comprehensively. The initial powders and their composites show the perovskite and spinel crystalline phases for the FE and FM fractions, respectively. After HPT, the particle sizes of the initial powders are decreased significantly. It is shown that the novel nanocomposite consists of exchange-interacting FE and FM phases and demonstrates improved magnetic and electrical properties in low fields at room temperature. A colossal increase in residual polarization with an increase in external high-pressure is found in new composite. The obtained results make it possible to consider the novel nanocomposite as a new functional material for its use both in electronic devices for monitoring ultra-high-pressure and in integrated circuits of high-speed computing nanosystems with low switching energy. The HPT method is a promising method for obtaining new heterophase multiferroic nanosystems.

*Keywords*: Multiferroic; Bismuth ferrite perovskite; Manganese-zinc ferrospinel; Nanocomposite; High pressure torsion; Magnetic properties; *P-E* loop

# 1. Introduction

Multiferroics are a new class of multifunctional materials where magnetic spin, electric dipole, and ferroelastic orderings coexist [1]. The coexistence of magnetostrictive and piezoelectric properties is the reason for the appearance of the magnetoelectric (ME) effect [2]. The increased interest in the designing and research of such new materials is due to their multifunctionality and practical application [3]. Based on the ME effect, supersensitive sensors of alternating and constant magnetic fields operate with more million times higher sensitivity than the Hall sensors [4]. The preservation of magnetostrictive properties in the microwave range is the basis for the creation of the latest technology for wireless energy transfer to miniature electronic devices [5]. Multilayer film heterostructures based on multiferroics have a high potential for use in high-speed storage devices with low energy consumption and high memory density [6]. Layered composite materials that suppress pyroelectric noise without averaging the useful signal are employed in the development of the latest high-sensitivity $\sim 10^{-12}$ T and low-frequency $\sim 10^{-2}$–$10^{3}$ Hz sensor technologies for non-invasive neurological interfaces in medicine [7].

Currently, the rapid development of modern technologies requires more and more new multiferroic materials including composites with improved characteristics and unique methods for their obtaining. The idea of creating a new composite is to find the phase and chemical composition of the heterophase system, which will respond to the improved functional properties for its practical application. It is advisable to consider the phase composition of the "ferroelectric-ferromagnetic", where the ferroelectric (FE) is La-modified $Bi_{0.9}La_{0.1}FeO_3$ (BLFO) multiferroic ferrite bismuth, and the ferromagnet (FM) is $Mn_{0.6}Zn_{0.3}Fe_{2.1}O_4$ (MZFO) manganese-zinc ferrospinel. Multiferroic BLFO has a high temperature of FE and FM orderings, as well as large electrical polarization values. The MZFO ferrospinel is a high-permeable magnetic ferrimagnet with a large initial magnetic permeability $\mu = 3000$, a high Curie temperature $T_C = 501$ K, a small coercivity $H_C = 0.25$ Oe, and the large saturation magnetization $M_S = 88$ emu/g [8]. It means that in small magnetic fields $H \sim 1$–100 Oe ($H > H_C$) at room temperature, the magnetic induction $B = \mu H$ may reach large values on the MZFO FM interface if MZFO particles are in a single-domain state. The critical size of the single-domain state $D_0$ for Mn-Zn ferrospinel is in the range $D_0 = 20$–30 nm [9]. For soft magnetized $MnFe_2O_4$ ferrospinel with close magnetic characteristics, the exchange length is $l_{ex} \approx 2$ nm [10]. If we can synthesize the BLFO–MZFO nanocomposite consisting of close-packed ultrafine particles with a size of $d < 20$ nm and with a width of interfacial boundaries less $l_{ex} \approx 2$ nm, we obtain a new multiferroic with a high probability of linear ME effect in the range of weak fields. In the new multiferroic, the FM phase of the single-domain MZFO regions located at a distance smaller than the FM exchange length will affect regions of the BLFO FE phase with a size of a smaller than a period of spin cycloid by an external magnetic induction increased in $\mu = 3000$ times. Owing to the influence of a strong FM phase, we will have a large resulting magnetic moment, which can be easily controlled by ME coupling.

Creating new FE materials based on nanocomposites is the most promising method for obtaining new multiferroics with a large value of ME coupling [2, 11]. Nowadays, most of the metal-oxide ceramic composites are obtained by mechanical mixing of powders with their sintering at the final stage [12]. However, this method is not suitable for our purpose, since the achievement of complete sealing at high sintering temperatures leads to a significant increase in grain size [13]. With the growth of grain, it may occur both the appearance of spin cycloids in the bismuth ferrite and a multi-domain state in the ferrospinel. In addition, at high sintering temperatures as a result of thermal dissociation, irreversible changes in the valence and spin state of manganese and iron are started with the degradation of the magnetic properties [14]. Therefore, a high-pressure torsion (HPT) method [15] for obtaining nanocomposite has been chosen, where a complete sealing of the heterophase system is accompanied by a statistically uniform distribution of phases in a bulk sample without changing the valence and spin state of magnetic ions [16].

The optimal relationship between the amount of FE and magnetic phases is necessary to find during designing and synthesizing a new composite in addition to the determination of the chemical and phase composition. Since the key phase is the FE phase, the number of BLFO particles should exceed the amount of MZFO particles. The quantitative composition of the BLFO–MZFO nanocomposite consisting of 80 wt.% BLFO and 20 wt.% MZFO corresponds to the model for the binary system [17], where the free space between the close-packed large BLFO particles of the multiferroic is filled with smaller MZFO particles of the ferromagnet. Necessary conditions for the implementation of such a model are the bigger size of the BLFO particles at least 3 times than the MZFO particles.

The purpose of our study is to create and study a new composite nanomaterial with a large ME coupling, which simultaneously will have a great value of electric polarization, and a large magnetic moment, as well as high (above room temperature) appearance of FE and FM orderings. The possibility of easy control of a large ME coupling of the multiferroic in the range of weak magnetic fields will allow us to obtain new functional material for its wide application in many scientific, technical, and medical applications.

## 2. Experimental section

*2.1. Sample preparation*

*2.1.1. The synthesis of the bismuth ferrite $Bi_{0.9}La_{0.1}FeO_3$*

The $Bi_{0.9}La_{0.1}FeO_3$ powder was synthesized using the nitrate pyrolysis method. The starting raw materials of $Bi(NO_3)_3 \cdot 5H_2O$ (99.99% metals basis, Sigma-Aldrich), $La(NO_3)_3 \cdot 6H_2O$ (99.9% metals basis, Sigma-Aldrich), and $Fe(NO_3)_3 \cdot 9H_2O$ (99.99% metals basis, Sigma-Aldrich) were dissolved separately in deionized water (75 mL) until a homogeneous solution. The stoichiometric mixture of the obtained solutions was evaporated to dryness in a water bath. The resulting dried powder was preheated at 600 °C for 2 h in the air to decompose metal nitrates:

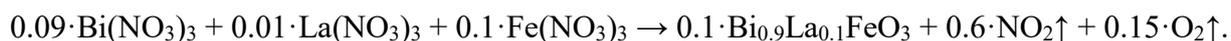

$0.09 \cdot Bi(NO_3)_3 + 0.01 \cdot La(NO_3)_3 + 0.1 \cdot Fe(NO_3)_3 \rightarrow 0.1 \cdot Bi_{0.9}La_{0.1}FeO_3 + 0.6 \cdot NO_2\uparrow + 0.15 \cdot O_2\uparrow.$

The obtaining light brown powder was ground in an agate mortar, placed in an alundum crucible, and calcined in a muffle furnace at 700 °C with isothermal holding for 5 hours.

*2.1.2. The synthesis of the manganese-zinc ferrite $Mn_{0.6}Zn_{0.3}Fe_{2.1}O_4$*

The $Mn_{0.6}Zn_{0.3}Fe_{2.1}O_4$ powder was prepared by the sol-gel auto-combustion method. The stoichiometric amounts of $Mn(NO_3)_2 \cdot 4H_2O$ (98%, Sigma-Aldrich), $Zn(NO_3)_2 \cdot 6H_2O$ (AR, 99%, Sigma-Aldrich), $Fe(NO_3)_3 \cdot 9H_2O$ (99.99% metals basis, Sigma-Aldrich) were dissolved in deionized water. The appropriate amount of fuel (citric acid monohydrate $H_3Cit \cdot H_2O$, ACS reagent, ≥ 99.0%) was added to the metal nitrate solution with continuous stirring. The molar ratio of the fuel to the total amount of metal ions (F/M) was set at 1. The final pH was adopted as 7 for better complexing of the metal ions. The pH was adjusted by adding 25 vol.% ammonia solution. The brown sols were heated until the xerogels have been formed. The gel was additionally dried at 120 °C and then rapidly heated at 200 °C to initiate the spontaneous combustion. Finally, the loose ferrite powders were obtained (see Fig.1).

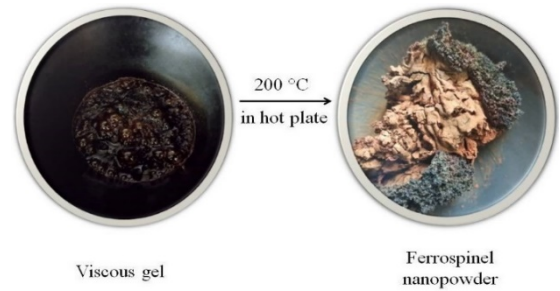

**Fig. 1**. Schematic illustration of ferrospinel nanopowder obtained by sol–gel auto-combustion method.

*2.1.3. BLFO–MZFO nanocomposite obtaining*

The BLFO–MZFO nanocomposite was prepared by the HPT from the powder mixture consisting of 80 wt.% BLFO and 20 wt.% MZFO. The obtained samples were a disk with a diameter of 10 mm and a thickness of 1 mm. This process was carried out in a custom-built HPT installation (W. Klement GmbH, Lang, Austria) with standard semi-constrained anvils (Fig. 2). The axial pressure 5 GPa was applied, and it was automatically maintained at this level during the experiment. The diameter of the anvil grooves was 10 mm and their total depth was 0.6 mm. The turning number of the anvils was 5. A rotational speed was 1 rpm. The experiment was carried out at ambient temperature.

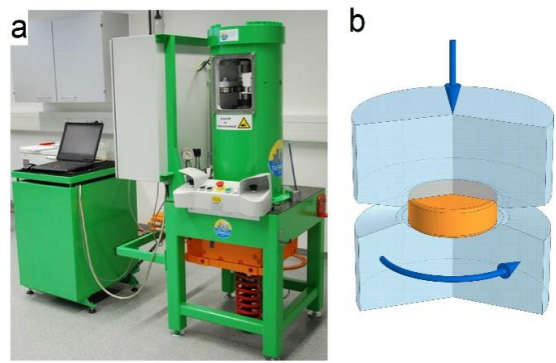

**Fig. 2**. The HPT device (a) and the HPT anvil circuit (b) for processing the BLFO–MZFO nanocomposite.

Additionally, the BLFO–MZFO composite was obtained under a pressure of 200 MPa but without HPT. The obtained samples were in a shape of disk with a diameter of 10 mm and a thickness of 1.5 mm. These samples were named as BLFO–MZFO composite without HPT.

**2.2. Methods**

The structure symmetry, phase composition, lattice parameters, and the size of the coherent scattering region were determined by the X-ray diffraction (XRD) method at room temperature using a Shimadzu Lab XRD 6000 diffractometer in CuKα-radiation ($\lambda Cu = 1.5406$ Å). The X-Ray tube was operated at a current of 30 mA and a voltage of 40 kV. The exposure time was 1 s and the measured angle region ($2\theta$) was from 5 to 70°. The scanning step was 0.02°. The structural refinement was performed with Rietveld analysis [18] using the FullProf software [19].

The morphology and particle size distribution of the BLFO and MZFO powders were determined using FEI MAGELLAN 400 Scanning Electron Microscope (SEM) and JEM-2200FS Transmission Electron Microscope (TEM). High-resolution TEM (HRTEM) with accelerating voltage 200 kV was employed to obtain information about the size and shape of the particles, as well as to determine an average interplanar distance using the fast Fourier transformation (FFT) approach. The powders for the TEM analysis were prepared by placing a drop of a diluted mixture of particles and acetone on a carbon-coated copper grid. An average particle size $D$ was obtained from analysis of SEM and TEM images within clear and defined particle borders using Nano Measure 1.2.5 software [20] during approximation of the experimental values of $D$ by different distribution functions. It should also be noted that obtaining clear images was quite complex due to magnetic attraction between the nanoparticles and their agglomeration in the high-temperature FM ordered MZFO nanopowder (see below Section 3.2). The chemical composition of all samples was performed by energy-dispersive X-ray spectroscopy (EDS) using an additional module of FEI MAGELLAN 400.

Magnetic measurements were performed using Quantum Design SQUID MPMS 3 in a wide temperature range from 2 to 900 K and in a magnetic field up to 30 kOe. Measurements of the temperature dependence of magnetization $M(T)$ in a field $H$ = 50 Oe were carried out in two regimes of zero-field cooling (ZFC) and field cooling (FC). The heating and cooling of samples were conducted with a constant rate equal to 1 K/min. The temperature relaxation time at the point of measurement was 2 min. The measurement of the field dependence of magnetization $M(H)$ at different temperatures was carried out with an increase in the magnetic field $H$ from 0 to 30 kOe with a step of $\Delta H$ = 100 Oe.

The FE properties based on the $P(E)$ hysteresis loops measured in the alternating electric field $E$ at room temperature were studied on a Precision Multiferroic II analyzer equipped by a charge-based magnetoelectric response tester using diamond anvil cell (DAC) [20]. The gasket with 300 $\mu$m in thickness of the pre-indented stainless steel was placed on the symmetric DAC with a culet diameter of 400 $\mu$m. The ground bulk samples were loaded into a cylindrical hole with a diameter of about 150 $\mu$m, together with a few ruby chips for pressure calibration. Two Pt strips were used as electrodes and placed on the surface of the sample. These measurements allow us to obtain $P(E)$ curves for applied maximum voltage 500 V and frequency of 50 Hz, within the electric field range from −60 kV·cm$^{-1}$ to +60 kV·cm$^{-1}$.

## 3. Results and Discussion

*3.1. Morphology and structural properties*

According to the XRD data (see Fig. 3), the initial BLFO and MZFO powders are fully single-phase. The XRD data indicates already a complete crystallization in the MZFO powder at 200 °C without additional annealing. The broadening of the diffraction maxima for the MZFO is due to the ultra-dispersity of the powder (see Fig. 3, *b*).

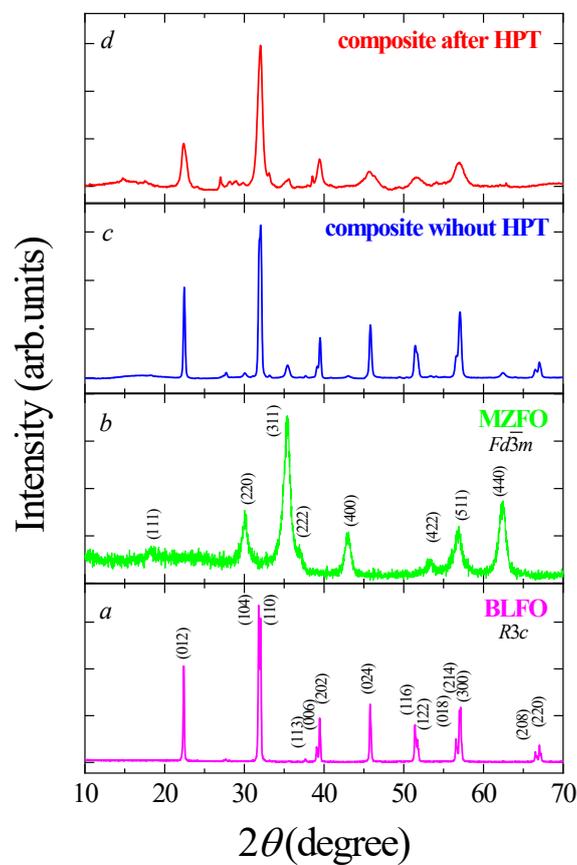

**Fig. 3**. XRD patterns of the BLFO (*a*) and MZFO (*b*) initial powders, as well as the BLFO–MZFO composite without HPT (*c*) and after HPT (*d*).

The crystal structure of the BLFO powder has a rhombohedral *R*3*c* distortion of perovskite structure (JCPDS No. 01-086-1519) with the $a = 5.57321$ Å and $c = 13.78585$ Å (see Supplementary Material (SM) 1) parameters of the unit cell. The MZFO powder has a cubic $Fd\bar{3}m$ spinel structure (JCPDS No. 96-230-0585) with the unit cell parameter of $a = 8.40195$ Å. The X-ray density determined from the XRD data is 8.217 and 5.238 g/cm$^3$ for the BLFO and MZFO powders, respectively. The main crystallographic parameters of the initial BLFO and MZFO powders refined by the Rietveld method are shown in Table S1.1.

The XRD pattern of the BLFO–MZFO composite without HPT (see Fig. 3 (*c*)) corresponds to a combination of the XRD patterns from the initial BLFO and MZFO powders (see Fig. 3 (*a*, *b*)). A distinctive feature of the XRD pattern for the BLFO–MZFO composite after HPT (see Fig. 3 (*d*)) is the X-ray line broadening due to a decrease in the crystallite size. An average crystallite size $D_{XRD}$ in powders and composites was calculated by the X-ray line broadening method using the Scherrer formula [21] in SM2. In the initial powders, the $D_{XRD}$ is 131 nm for the BLFO powder and 15 nm for the MZFO powder (see Table 1). In a two-phase BLFO–MZFO composite without HTP, the size of the BLFO fraction decreased from 131 to 103 nm, while the size of the MZFO fraction was the same. In the BLFO–MZFO composite after HPT, there was a severe decrease in the size of the multiferroic BLFO fraction from 131 to 14 nm and a slight reduction in the size of the FM MZFO fraction from 15 to 12 nm. Such a decrease in the size of BLFO and MZFO fractions is the result of crushing the large BLFO particles with a hardness of ~ 2.5 GPa [22] by the small MZFO particles with a higher hardness of ~ 5.5 GPa [23] during plastic deformation of the HPT.

**Table 1**

Comparative analysis of XRD, TEM, and SEM data for the particle size of the initial BLFO and MZFO powders, and BLFO–MZFO composites without and after HPT.

| Sample | | $D_{XRD}$ (nm) | $D_{TEM}$ (nm) | $D_{SEM}$ (nm) |
|---|---|---|---|---|
| BLFO powder | | 131 | 158 | 370 |
| MZFO powder | | 15 | 15 | – |
| BLFO–MZFO composite without HPT | BLFO fraction | 103 | 75 (70%) 143 (9%) | – |
| | MZFO fraction | 15 | 15 (21%) | – |
| BLFO–MZFO composite after HPT | BLFO fraction | 14 | 13 | – |
| | MZFO fraction | 12 | 12 | – |

According to TEM data (see Fig. 4 and SM2), the initial BLFO and MZFO powders consist of spherical-like particles with a size of $D_{TEM}$ = 158 and 15 nm, respectively, which are consistent with the $D_{XRD}$ values (see Table 1). Moreover, the BLFO and MZFO samples demonstrate well crystalline structure according to the selected area electron diffraction (SAED) pattern (see upper inserts of Fig. 4) with an interplane distance of 0.382 nm (012) for the perovskite phase and 0.483 nm (111) for the spinel phase (see bottom inserts of Fig. 4). Obtained data confirm the possibility of the spinel structure crystallization at low temperatures of ~ 200 °C without additional synthesis at high temperature ~ 1000 °C. This is the important result for nanotechnology and nanomagnetism since additional annealing at higher temperatures can lead both to significant grain growth with standing beyond the nanoscale phenomena [24], and a change in the ion valence states with declining the functional properties of the magnetic materials [25].

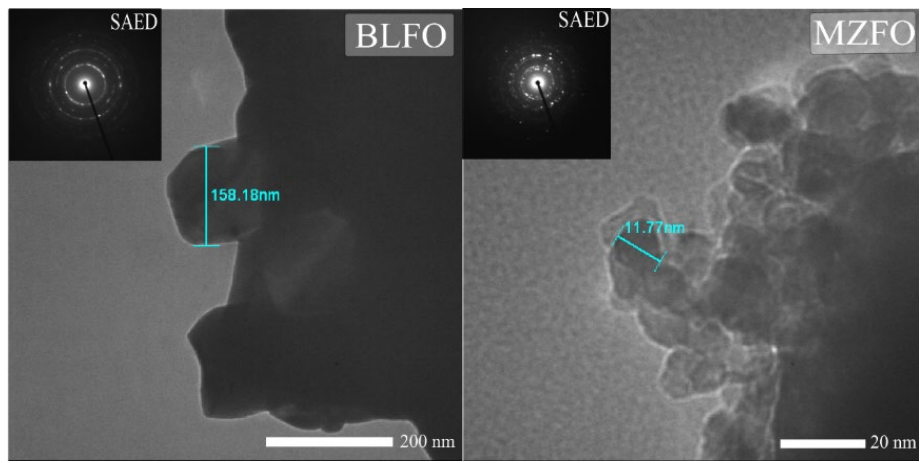

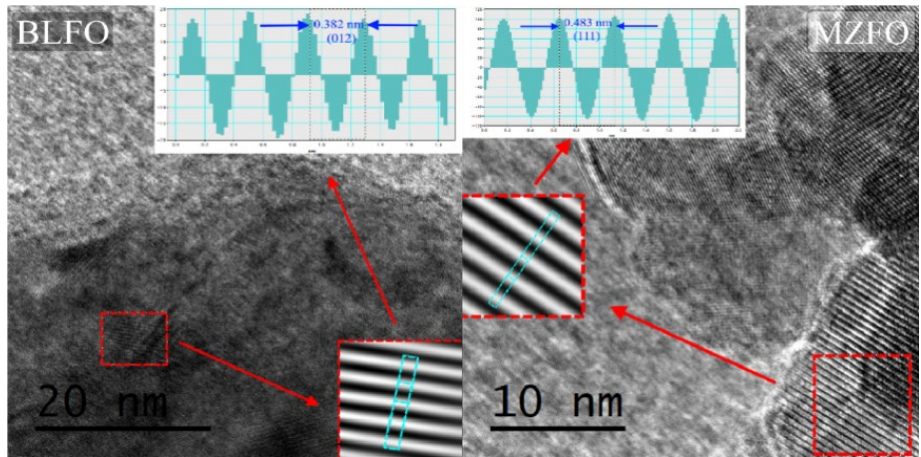

**Fig. 4**. The TEM and HRTEM images for the initial BLFO and MZFO powders. The inserts show SAED and the FFT of the HRTEM images with the lattice plane intensity profile.

Using TEM data (see Fig. 5, top), it can be seen that the BLFO–MZFO composite without HPT corresponds to a mixture of BLFO and MZFO powders, where the size of the MZFO particles maintains, and the size of the BLFO particles decreased by ~ 2 times under a compacting pressure of 200 MPa. The percentage of particles with $D_{BLFO}$ = 75 nm (89%) and 143 nm (11%) have been determined by approximating the $n(D)$ dispersion by the Gaussian function, considering the normalization condition of $\int_0^{+\infty} n(D)dD = 1$ (see Fig. SM3).

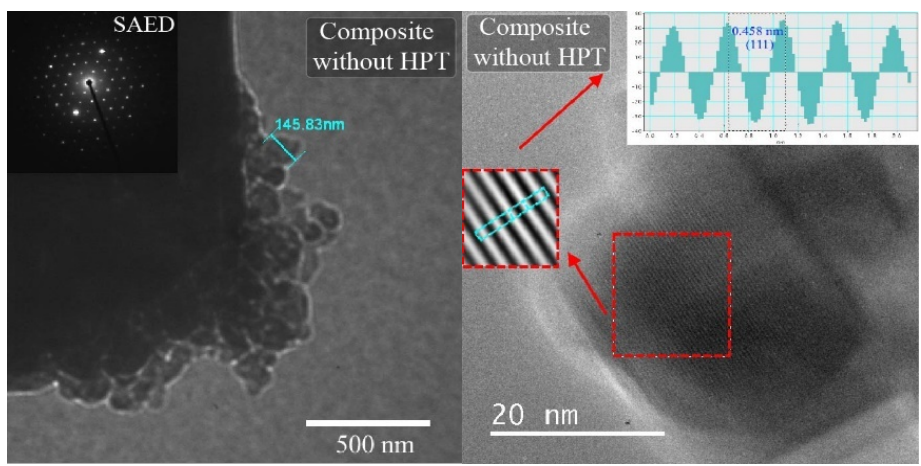

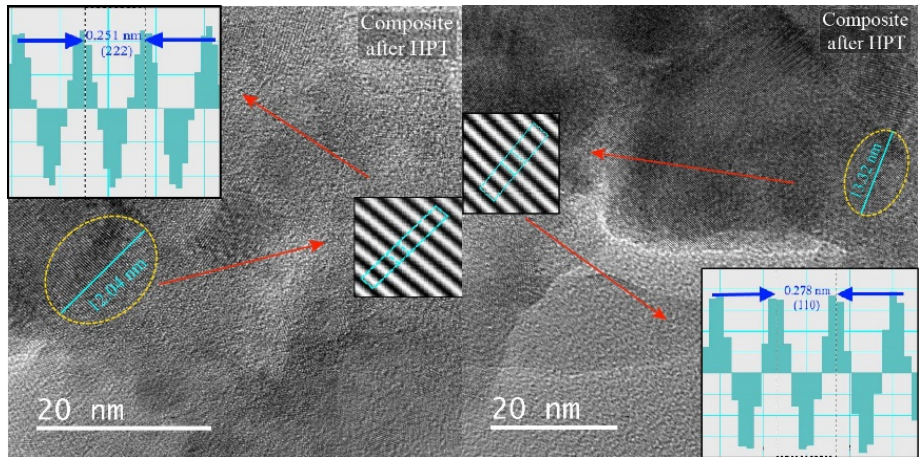

**Fig. 5**. The TEM and HRTEM images for the BLFO – MZFO composite without HPT ($p = 0$, top) and after HPT ($p = 5$ GPa, bottom). The inserts show SAED, and the fast Fourier transformation (FFT) of the HRTEM images with the lattice plane intensity profile.

However, the significant changes in the BLFO–MZFO composite after HPT (see Fig. 5, bottom) have been occurred. Firstly, the HRTEM images show a more than 10-times decrease in the size of the BLFO fraction from $D_{TEM} = 158$ to 13 nm, which is in good agreement with XRD data (see Table 1). Secondly, the BLFO and MZFO fractions were compacted very tightly that led to weakly distinguishable shades in TEM images for the multiferroic and FM phases. For the accurate determination of the size of the BLFO and MZFO fractions, the differences in the values of interplanar distance according to the HRTEM images were used. Analysis of HRTEM images made it possible to estimate the width of the interphase boundaries of ~ 1 nm. In the next section, a comparison of the width of the interphase boundaries with the FM exchange length $l_{ex}$ will play important role in an increasing the ME coupling for a composite of close-packed multiferroic and FM fractions as a result of induced ferromagnetism into FE region from the FM region.

According to the SEM data (see Fig. 6 and SM3), the BLFO–MZFO composite without HPT (see Fig. 6 (*e*)) is a combination of the large BLFO with $D_{SEM} = 370$ nm (see Fig. 6 (*a*)) and small MZFO (see Fig. 6 (*c*)) particles. The large size of $D_{SEM}$ compared to $D_{XRD}$ and $D_{TEM}$ is due to the adhesion [26]. At the same time, the BLFO–MZFO composite after HPT is well compacted and has a homogeneous microstructure without clear boundaries (see Fig. 6 (*g*)).

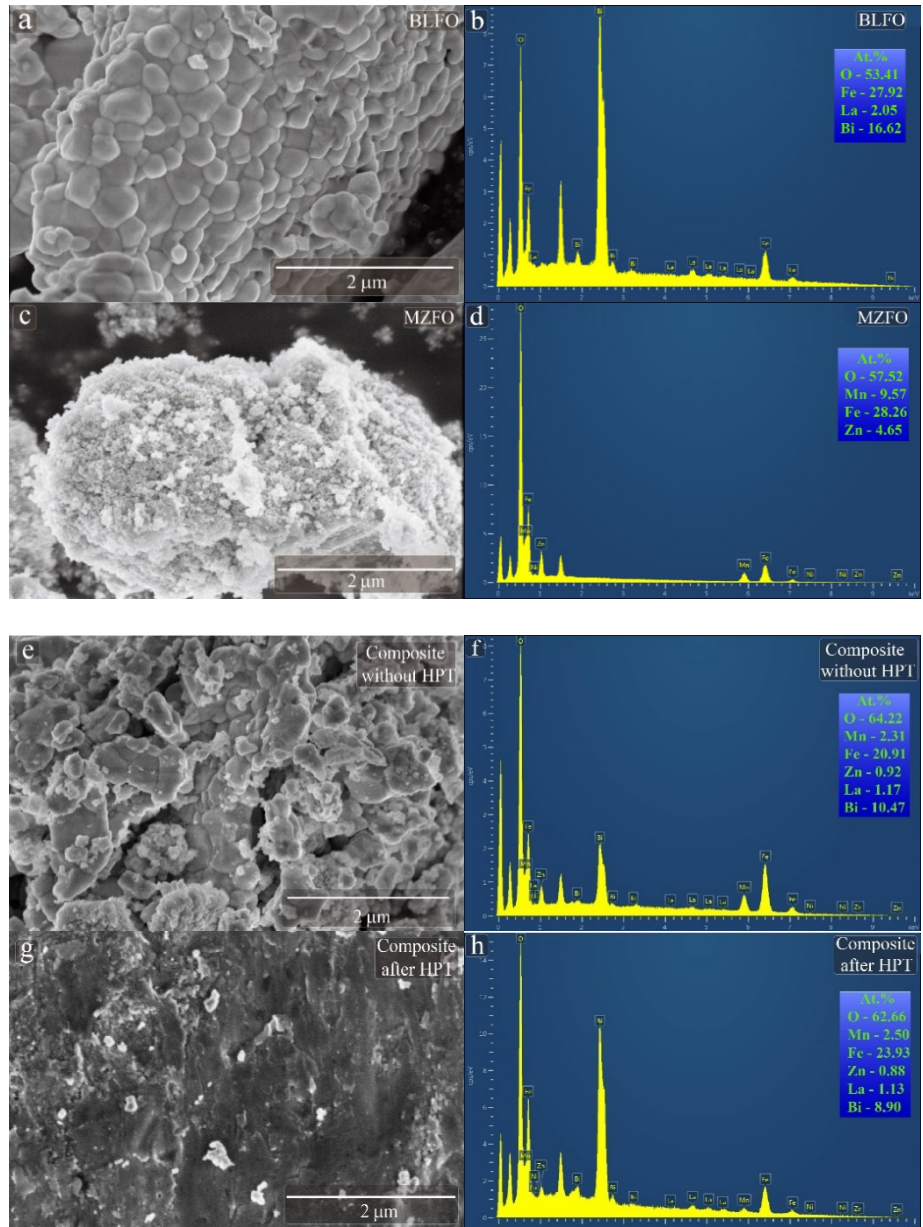

**Fig. 6**. The SEM and EDS data for the initial BLFO and MZFO powders, as well as the BLFO–MZFO composites without and after HPT.

According to the EDS data, the chemical composition of the BLFO and MZFO powders, as well as both BLFO–MZFO composites, is confirmed, which corresponds approximately to the stoichiometric compositions of $Bi_{0.9}La_{0.1}FeO_3$ and $Mn_{0.6}Zn_{0.3}Fe_{2.1}O_4$ and their combinations (see Fig. 6 and SM4).

Thus, using torsion under high-pressure is an effective and promising method for obtaining heterophase nanostructures for novel type 2 multiferroics consisting of close-packed, ultra-dispersed, and uniformly distributed multiferroic and FM phases. Torsion under high-pressure results in (i) a statistical uniform distribution of fractions in the sample; (ii) reducing the size of phases due to the crushing of the particles by others with a higher microhardness that can be controlled by the turning number; (iii) decreasing the width of the interphase boundaries until to several unit cell periods at the close-packed phases.

### 3.2. Magnetic properties

The temperature dependences of the magnetization $M_{ZFC}(T)$ and $M_{FC}(T)$ for the initial BLFO powder indicate the AFM nature of the interactions (see Fig. SM5.1) [7a]. High values of coercivity $H_C \approx 2.2-2.4$ kOe and remanent magnetization $M_r \approx 0.0035$ emu/g in the range of room temperatures (see Table 2) compared to $H_C \approx 0.003$ kOe and $M_r \approx 0.0005$ emu/g for the model multiferroic $BiFeO_3$ with cycloidal AFM [27] mean the destruction of the spin cycloid in the AFM iron sublattice by lanthanum ions. The appearance of hysteresis in the field dependences of $M(H)$ (see Fig. SM5.2) is associated with the presence of weak FM in the BLFO powder [28].

**Table 2**

The coercivity $H_C$, remanent magnetization $M_r$, and saturation magnetization $M_S$ for the initial BLFO and MZFO powders, as well as BLFO–MZFO composite without and after HPT

| Sample | BLFO | | | MZFO | | | composite without HPT | | | | composite after HPT | | | |
|---|---|---|---|---|---|---|---|---|---|---|---|---|---|---|
| $T$ (K) | 2 | 300 | 400 | 2 | 300 | 400 | 2 | 77 | 300 | 400 | 2 | 77 | 300 | 400 |
| $H_C$ (Oe) | 1150 | 2275 | 2425 | 365 | 55 | 82 | 352 | 135 | 7 | 23 | 337 | 127 | 35 | 34 |
| $M_S$ | – | – | – | 88 | 59 | 47 | 16 | 15 | 11 | 8 | 9 | 8 | 3 | 0.4 |
| (emu/g) | | | | | | | | | | | | | | |
| $M_r$ (emu/g) | 0.024 | 0.032 | 0.036 | 26 | 2 | 3 | 5 | 2 | 0.2 | 0.4 | 3 | 2 | 0.4 | 0.04 |

The temperature and field dependences of the $M_{ZFC}(T)$, $M_{FC}(T)$ and $M(H)$ for the initial MZFO powder (see SM5) show its highly permeable soft ferromagnet nature (see Table 2). The Curie temperature $T_C$ = 518 K (see the inset of Fig. SM5.1) is above the room temperature and coincides with $T_C$ for Mn-Zn ferrospinels of industrial grades 3000HMC (USSR), 3C91 (Germany), ML27D (Japan), which have the same stoichiometric composition and a large initial magnetic permeability $\mu$ = 3000 [8, 25, 29]. As a temperature increases, the magnetic parameters decrease from $H_C$ = 365 Oe, $M_S$ = 88 emu/g, and $M_r$ = 26 emu/g at $T$ = 2 K to $H_C \approx$ 50–80 Oe, $M_S \approx$ 50 emu/g, and $M_r$ = 2–3 emu/g for the entire range of room temperatures from 300 to 400 K. Such magnetic characteristics make it possible to easily control the magnetic properties of FM high-permeability MZFO powder with $\mu$ = 3000 by relatively small magnetic fields $H \sim$ 50–100 Oe.

Fig. 7 shows the $M_{ZFC}(T)$, $M_{FC}(T)$ and $M(H)$ dependences for both BLFO–MZFO composites without and after HPT. For the BLFO–MZFO composite without HPT, with the increase in the temperature, the coercivity and magnetization decrease (see Table 2). The $T_C$ = 518 K has not changed. The non-monotonic behavior of the $H_C(T)$ at the room temperature is due to the presence of two uncompensated AFM sublattices in the ferrospinel with different temperature dependences of magnetization [30]. Analyzing the magnetic properties, it was found that the BLFO–MZFO composite without HPT did not exhibit individual features and its magnetic characteristics correspond to the additive contribution of magnetic properties from 80 wt.% BLFO and 20 wt.% MZFO.

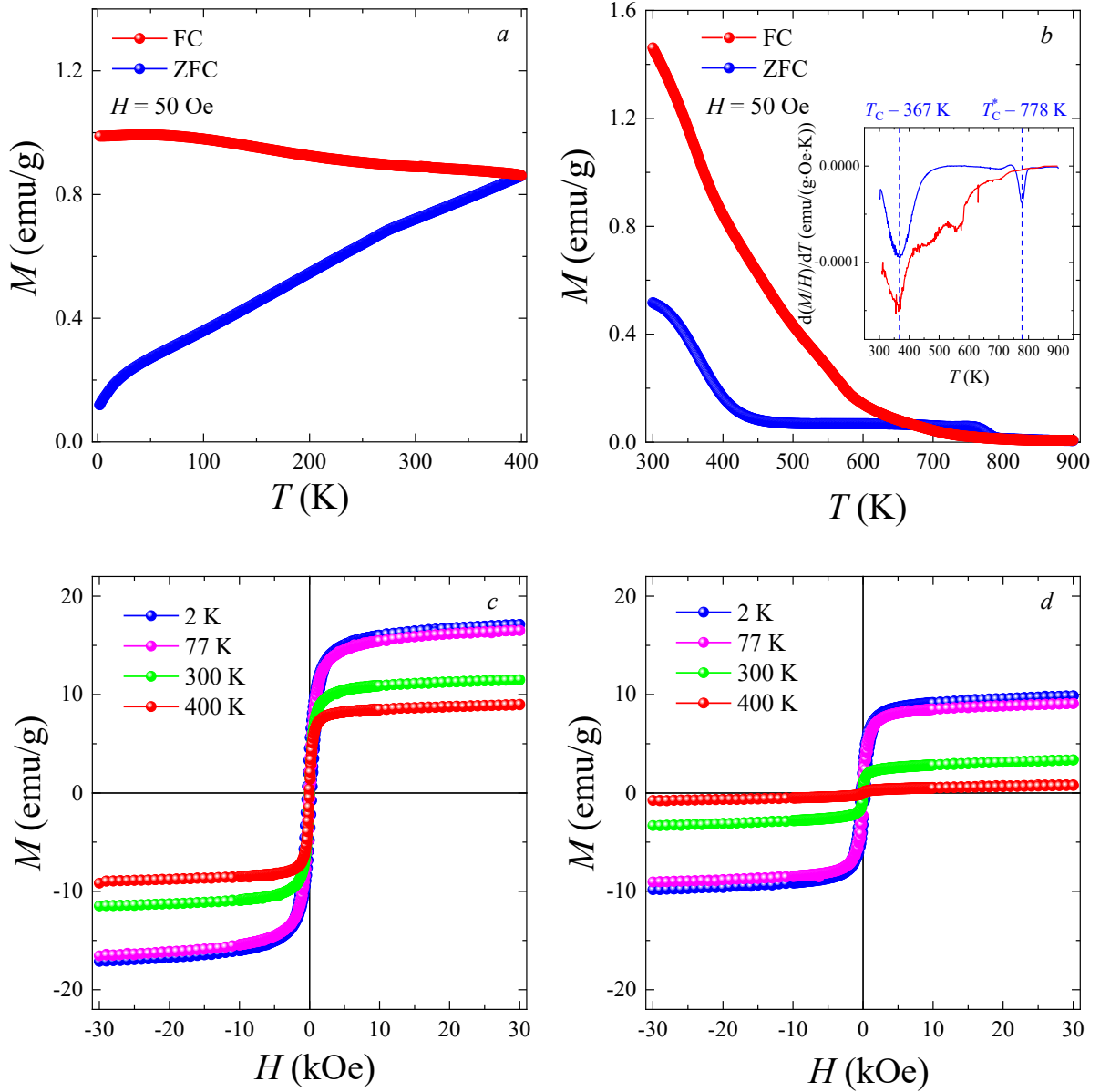

**Fig. 7**. The $M_{ZFC}(T)$ and $M_{FC}(T)$ temperature dependencies (a,b) and $M(H)$ field dependencies (c,d) of the magnetization for the composite without (*a* and *c*) and after HPT (*b* and *d*). The inset shows the Curie temperature $T_C$.

Compared to the composite without HPT, the BLFO–MZFO composite after HPT showed the properties of new multiferroic material (see Fig.7 and Table 2): (i) the $T_C$ has significantly decreased from 518 to 367 K, that is in the room temperature range; (ii) the behavior of the $H_C(T)$ became monotonic for the entire room temperature range with a stabilized valued of $H_C \approx 35$ Oe;

(iii) a new FM order appeared at $T_C^* = 778$ K; (iv) an anomalous hysteresis appeared between the $M_{ZFC}(T)$ and $M_{FC}(T)$ dependences, where the ZFC curve is above the FC curve in the temperature range from 671 to 795 K; (v) in the $M_{FC}(T)$ in contrast to the $M_{ZFC}(T)$ dependence, there are no signs of FM ordering at $T_C^*$; the residual magnetization reduces to $M_r \approx 0.04$–0.4 emu/g without a significant reduction in the $M_S \approx 3$ emu/g at the room temperature. It is necessary to note that the magnetization at $H = 50$ Oe increases 250 times from $M = 0.002$ to 0.517 emu/g and the coercivity decreases 65 times from $H_C = 2275$ to 35 Oe in the BLFO–MZFO composite after HPT compared with the BLFO at $T = 300$ K. Moreover, the calculated value of the induction of the magnetic field $B = \mu H$ can take large values of ~ 15T on the border of the MZFO fraction with $\mu = 3000$ and $H = 50$ Oe in the new composite. However, the $B$ cannot exceed the saturation induction $B_S$ when all magnetic moments of iron are aligned along with the $H$. For the MZFO powder with the $M_S = 88$ emu/g and a density of 5.238 g/cm³ (see Tables 1 and 2), the $B_S = 4\pi \cdot 10^{-4} M_S$ (emu/cm³) [31] equals 0.58 T. Such $B_S$ value is well consistent with the experimental value of $B_S = 0.525$ T for the manganese ferrospinel, which has an uncompensated total magnetic moment $4.2\mu_B$ in the ferrimagnetic iron sublattice [32]. The appearance of the $B = 0.58$ T in the composite means that the FM MZFO fractions with $\mu = 3000$ even in a small $H \sim 50$ Oe will induce a large inner $H_{int} \sim 5800$ Oe acting on the FE BLFO fraction.

The existence of interactions between FM MZFO and FE BLFO fractions in the BLFO–MZFO composite after HPT indicates anomalous temperature hysteresis on the $M_{ZFC}(T)$ and $M_{FC}(T)$ dependencies within the 671–795 K (see Fig. 7). Anomalous behavior of the ZFC curve above the FC curve is due to an additional FM contribution to magnetization when the composite is heated from the low-temperature FM region and the absence of this contribution when it is cooling from its high-temperature PM region. The MZFO fraction is the FM below $T_C$ = 367 K. The BLFO fraction is the AFM below $T_N$ = 660 K, FE below the temperature $T_C$ = 950 K, and paraelectric (PE) above this temperature [33]. The phase state of the BLFO–MZFO composite after HPT can be represented as: FM/(AFM+FE) ← ($T_C$ = 367 K) → PM/(AFM+FE) ← ($T_N$ = 660 K) → PM/(PM+FE) ← ($T_C$ = 950 K) → PM/(PM+PE).

The anomalous appearance of the FM contribution in the PM/(PM+FE) phase (see the inset in Fig. 7) is a consequence of a metamagnetic phase transition [7a, 34] in the FM/(AFM+FE) phase. The metamagnetic phase transition is a reorientation process with the "order-order" type phase transition induced by a field, which changes the type of magnetic ordering [35] or an angle between magnetic sublattices in the magnet [36]. The metamagnetic phase transition can be induced not only by external $H$ but also internal $H_{int}$ [36]. As was shown [37], an additional FM contribution to the magnetization of nanoscale AFM perovskite $BiFeO_3$ with a weak Dzyaloshinsky-Moriya interaction can occur as a result of a spin-reorientation transition induced by the magnetic field. Moreover, this FM contribution depends on both the magnitude of the $H$ and the stress-strain. The FM MZFO fractions in a weak field ~ 50 Oe induce large internal fields $H_{int}$ ~ 5800 Oe. Since the exchange length $l_{ex} \approx 6$ nm in the MZFO (see SM6) exceeds the width of the interfacial boundaries of ~ 1 nm and is comparable to the size of ~ 14 nm of the BLFO fractions, we can conclude about the existence of exchange interactions between BLFO and MZFO fractions. The presence of exchange interactions, the large $H_{int}$, and the high stress-strain after HPT are the necessary conditions for the spin-reorientation transition in the FM/(AFM+FE) phase [38], which is the result of an additional FM contribution to the $M_{ZFC}(T)$ dependence. Internal stresses and magnetostriction stabilize FM interactions [39] up to the $T_C^*$ = 778 K, above which the BLFO fraction goes into the PM state. An increase in the temperature range of anomalous hysteresis to 795 K is due to spin fluctuations in the frustrated magnetic subsystem of the bismuth ferrite [40]. When the composite is cooled, an additional FM contribution to the $M_{FC}(T)$ is not observed due to the lack of the $H_{int}$, since the MZFO fractions are in the PM state. The presence of exchange interactions and the irreversible nature of the additional FM contribution to the magnetization is the reason for the appearance of anomalous temperature hysteresis in the temperature range from 671 to 795 K.

The appearance of the exchange-related state between the "strong" FM subsystem and FE phase in the novel composite should lead to a significant increase in the ME coupling in the range of weak magnetic fields. For this purpose, it is necessary to analyze the hysteresis $P(E)$ curves electronic polarization $P$ in the electric field $E$ and establish that the HPT does not impair the FE properties of the initial BLFO powder.

### 3.3. *P-E* properties

The ferroelectric hysteresis $P(E)$ loops of the initial BLFO powder, as well as the BLFO–MZFO composite without and after HPT at the different $E$ are shown in Fig. 8. All samples demonstrate the FE behavior for the different applied voltages with a stepwise increase in $E$ up to a maximum value of $E_{max}$ = 18 kV/cm. The choice of $E_{max}$ was not accidentally because of the need to perform the equality of $w_E = w_H$. According to this equality, the energy density of the electric field $w_E = \frac{1}{2}\varepsilon\varepsilon_0 E_{max}^2$ ($\varepsilon_0$ = 8.85·10$^{-12}$ F/m is a vacuum permittivity) [41] has to coincide with the energy density of the magnetic field $w_H = \frac{1}{2}BH$ in the field $H$ = 50 Oe from the magnetic measurements. A dielectric constant $\varepsilon$ equals 80 for the BLFO multiferroic in the electromagnetic field with a frequency of 50 Hz [42]. In the MZFO ferromagnet at the $H$ = 50 Oe (3980 A/m), the induction $B$ is 0.6 T (see 3.2. Magnetic properties). With such values of the ε, B, and H, the equality of $w_E = w_H$ will be performed in the $E$ = 18.4 kV/cm. Additional measurements of frequency dependences of polarization $P$ confirm the absence of a leakage current at a frequency of 50 Hz [43], which indicates the correct formulation of the experiment and the definition of $E_{max}$ = 18 kV/cm.

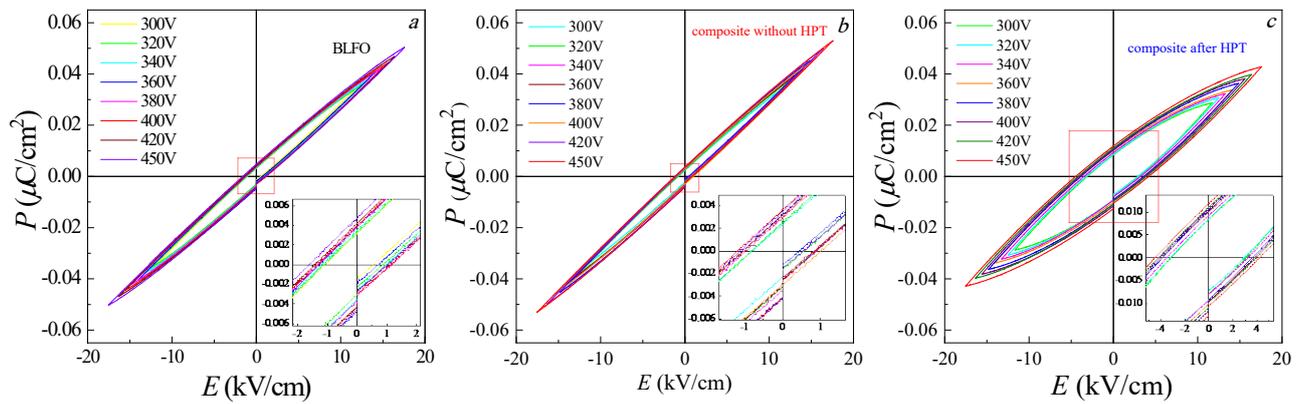

**Fig. 8**. Ferroelectric hysteresis loops for the initial BLFO powder (*a*), as well as the composite without (*b*) and after HPT (*c*).

The shape of *P-E* loops indicates that the $E_{max}$ has not reached the saturation field. The maximum polarization of $P_{max}$ at $E_{max}$, the residual polarization $P_r$, and the coercive field $E_C$ with a stepwise increase in the maximum measuring field from 11.7 to 18.0 kV/cm were defined (see Fig. 9(*a*)). In all samples, the linear course of the $P_r(E)$ was found for the entire range of *E*. In the BLFO powder, the $P_{max}$, $P_r$, and $E_C$ are 0.052 $\mu C/cm^2$, 0.005 $\mu C/cm^2$, and 1.6 kV/cm, respectively. In the BLFO–MZFO composite without HPT, a slight decrease in the $P_r$ to 0.004 $\mu C/cm^2$ and the $E_C$ to 1.3 kV/cm is due to a decrease in the FE content of the BLFO phase to 80 wt.%. This conclusion is confirmed by parallel displacement along the ordinate axis of the $P_r(E)$ dependence for the composite without HPT (see Fig. 9 (*a*)). The increase in the $P_{max}$ from 0.052 to 0.055 $\mu C/cm^2$ is caused by a decrease in the average particle size of the BLFO in the composite from 131 to 103 nm (see Table 1). It leads to an increase in the total surface of the FE phase and the accumulation of a weak-connected charge on the boundaries in the low-frequency range [44]. The comparative analysis of *P-E* loops between the initial BLFO powder and BLFO–MZFO composite without HPT (see Fig. 8 (*a* and *b*)) allows us to conclude that the FE properties of the composite are due to the 80 wt.% BLFO fraction and they weakly differ from the properties of the BLFO.

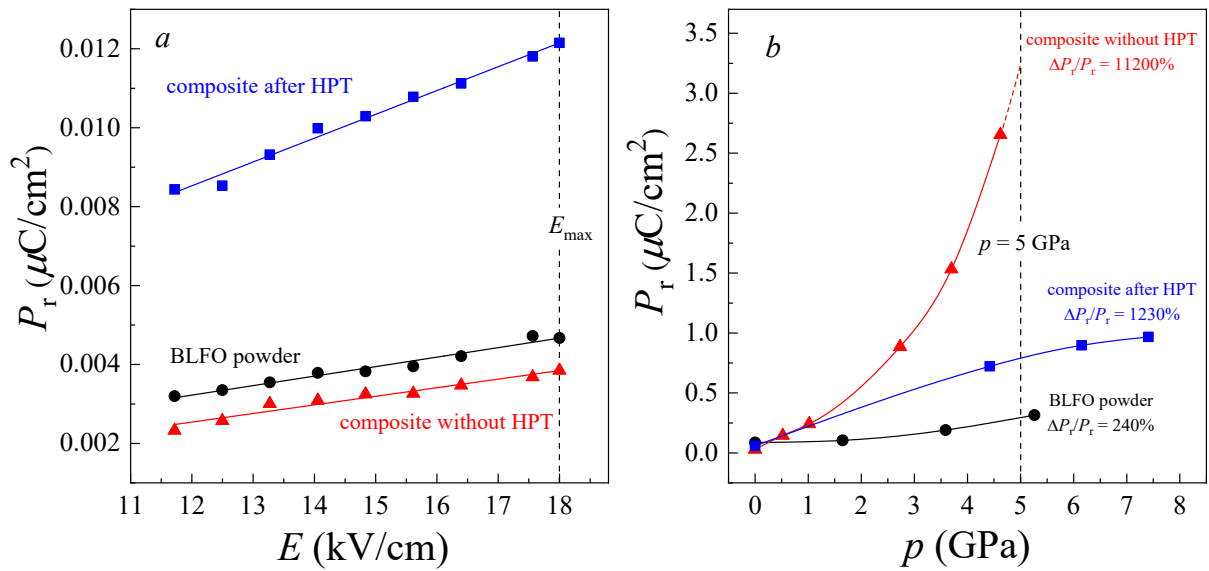

**Fig. 9**. The electric field (*a*) and pressure (*b*) dependences of the residual polarization for the BLFO powder and BLFO–MZFO composites without and after HPT. The $\Delta P_r/P_r$ is a relative change in the pressure-induced residual polarization under 5 GPa.

In the BLFO–MZFO composite after HPT, there are significant improvements in the FE properties compared to the BLFO powder. Even without considering the decrease in the amount of FE phase in this composite, an increase of 2.6 times in the residual polarization to $P_r = 0.012\ \mu C/cm^2$ and an increase of 2.9 times in the coercive field to $E_C = 4.6$ kV/cm are observed. Such a broadening of the hysteresis region is due to an increase in the ME coupling. The magnetic field induced by the alternating electric field gives an additional contribution to the polarization of the multiferroic in the presence of interaction between FE and FM phases [45]. Reducing the $P_{max}$ by ~ 15% to 0.043 $\mu C/cm^2$ is due to the presence of 20 wt.% non-ferroelectric magnetic MZFO fraction in the composite.

Evaluation of the energy density of the electric field $w_E$ for the new composite at room temperature allows us to conclude that a small alternating magnetic field $H \sim 50$ Oe can induce relatively large electric fields ~ 18 kV/cm with an energy ~ $3 \cdot 10^{-21}$ J in the FE fractions with a size of ~ 14 nm and with the dielectric constant $\varepsilon \sim 80$. With such properties, the new nanocomposite can become a promising functional material during the creation of high-performance artificial intelligence systems [46]. Such a nanocomposite can dramatically solve the problem of cooling and removing heat in the integrated circuits of high-speed computing systems associated with a decrease in switching energy for nanoscale devices [47].

The dependences of the effect of high-pressure $p$ on the residual polarization $P_r$ are shown in Fig. 9 (*b*). From the analysis of the $P_r(p)$ dependences, it can be concluded that there is a significant difference in the effect of $p$ on the FE properties of the initial BLFO powder and composites. In the BLFO, as $p$ increases from 0 to 5.26 GPa the residual polarization $P_r$ increases by ~ 4 times to 0.315 $\mu C/cm^2$ with a polarization sensitivity coefficient to 0.043 ($\mu C \cdot cm^{-2}$)/GPa. In the BLFO–MZFO composite without HPT, there is a colossal increase in the $P_r$ ~ 92 times up to 2.654 $\mu C/cm^2$ with an increase in $p$ from 0 up to 4.62 GPa with 0.568 ($\mu C \cdot cm^{-2}$)/GPa. In the BLFO–MZFO composite after HPT, the residual polarization increases by ~ 16 times to $P_r = 0.968$ $\mu C/cm^2$ under $p = 7.41$ GPa with 0.123 ($\mu C \cdot cm^{-2}$)/GPa.

For a quantitative assessment of the effect of high-pressure on FE, we have first introduced the relative change of the pressure-induced polarization coefficient $\Delta P_r/P_r = [P_r(p) - P_r(0)]/P_r(0) \cdot 100\%$ at the same pressure $p = 5$ GPa. This coefficient is $\Delta P_r/P_r = 237$ % for the BLFO powder, $\Delta P/P = 11200$ % for the composite without HPT, and $\Delta P_r/P_r = 1230$ % for the composite after HPT (see Fig. 9 (b)). The large changes in the $\Delta P_r/P_r = 1230$ % for the BLFO–MZFO composite after HPT compared to $\Delta P_r/P_r = 237$ % for the BLFO powder can be explained by the manifestation of anisotropic nature of piezoelectric properties [48] in the close-packed structure of the composite. However, the colossal change $\Delta P_r/P_r = 11200$ % in the BLFO–MZFO composite without HPT requires additional analysis.

An electrical polarization in the $BiFeO_3$ with a rhombohedral $R3c$ distortion of an elementary cell occurs as a result of an anti-phase rotation of the $FeO_6$ octahedra with a displacement of cations of iron and bismuth in one direction and anions of oxygen in another direction [49]. The result of such a local breaking symmetry of the crystal is the appearance of the vector of electrical polarization $P$ along the axis [111]. Applying external high-pressure enhances the deformation of octahedra, changes the angles and the lengths of metal-oxygen bonds, affects the structural properties, and leads to the emergence of new phase transformations [50].

In the range of external high-pressure from 1.25 to 4.5 GPa for the $BiFeO_3$, several phase transformations associated with the appearance and coexistence of polar $R3c$ rhombohedral (FE), nonpolar $P$nma orthorhombic (PE, $GdFeO_3$-type), antipolar $P$bam orthorhombic (AFE, $PbZrO_3$-type), and nonpolar $I$bmm orthorhombic (PE, nonstandard setting of $I$mma) phases occurs [51]. In addition to the polarization vector $P$ along the direction [111] in the FE $R3c$ phase, the vector $P$ along the direction $[110/\bar{1}\bar{1}0]_c$ in the AFE $P_{bam}$ phase is added [52]. During the coexistence of two different FE phases, a tremendous gain of piezoelectric properties occurs as a result of not only polarization rotation, but also polarization extension near the morphotropic "polar-antipolar" boundary [53]. Pressure-induced phase transformations in the $Bi_{0.9}La_{0.1}FeO_3$ will occur at lower ~

0.5 GPa external pressure because the chemical pressure during replacing bismuth by 1 mol.% of lanthanum in the $Bi_{1-x}La_xFeO_3$ is approximately equivalent to applying 0.05 GPa hydrostatic pressure on $BiFeO_3$ [54]. In the BLFO–MZFO composite without HPT, the colossal increase in polarization $\Delta P_r/P_r$ starts at ~ 1 GPa (see Fig. 9 (*b*)) and coincides with an appearance of the pressure-induced AFE $P_{bam}$ phase [54]. Such a colossal growth of $\Delta P_r/P_r$ is due to an additional contribution to polarization, both from the new AFE $P_{bam}$ phase and from expanding the polarization area near the border between the FE $R3c$ rhombohedral and AFE $P_{bam}$ orthorhombic phases. Additional contribution to pressure-induced grow of $\Delta P_r/P_r$ can give pinning on interfacial "FE/AFE" ($R3c/P_{bam}$), "FE/PE" ($R3c/P_{nma}$, $R3c/I_{bmm}$), and "AFE/PE" ($P_{bam}/P_{nma}$, $P_{bam}/I_{bmm}$) boundaries [7a].

Large values of pressure-induced relative changes in the polarization $\Delta P_r/P_r$ for the new composite are of interest and are of great practical importance because such a functional material can be used to create high-pressure sensors. The colossal values of $\Delta P_r/P_r$ = 11200 % at 5 GPa make it possible to use the BLFO–MZFO composite without HPT as a sensing element in ultrasensitive high-pressure sensors. For creating electronic devices to monitor high-pressure, where the presence of ME coupling in the sensitive element is required, it is advisable to use the BLFO–MZFO composite after HPT with $\Delta P_r/P_r$ = 1230 % at 5 GPa and with an almost linear dependence of $\Delta P_r/P_r$ on the pressure in a wide range from 0 to 7.5 GPa.

## 4. Conclusion

Both new "multiferroic-ferromagnetic" composite materials based on the lanthanum-modified bismuth ferrite (BLFO) and manganese-zinc ferrite spinel (MZFO) have been obtained without and after the high-pressure torsion (HPT) method. The comparative characteristics of their functional properties between each other and with initial BLFO and MZFO powders have been carried out. It has been shown that the nanocomposite after HPT is the most close-packed heterophase system consisting of statistically uniformly distributed ultrafine ferroelectric (FE) BLFO fractions of perovskite with a size of 14 nm and ferromagnetic (FM) MZFO fractions of spinel with a size of 12 nm.

It has been found that the BLFO–MZFO nanocomposite after HPT has the unique physical properties of new material with strong magnetoelectric coupling: (i) an appearance of Curie temperature $T_C$ = 367 K near the room temperature; (ii) the monotonic change in the coercivity $H_C$ vs. $T$ and its small value $H_C \approx 35$ Oe at $T = 300$ K; (iii) a decrease in the residual magnetization to $M_r \approx 0.04$–0.4 emu/g; (iv) an increase of 2.6 times in the residual polarization to $P_r = 0.012$ $\mu$C/cm$^2$ and an increase of 2.9 times in the field to $E_C$ = 4.6 kV/cm. It has been shown that a small alternating magnetic field of ~ 50 Oe has to induce relatively high electric fields of ~ 18 kV/cm with an energy of ~ 3·10$^{-21}$ J in the FE fractions of the nanocomposite after HPT.

Additionally, for the first time, the FE high-pressure studies of the composites without and after HPT, as well as BLFO powder have been conducted. As turned out, the composite without HPT has demonstrated the highest increase in the residual polarization under $p$ = 5 GPa.

Thus, we have obtained type 2 multiferroic composites with strong magnetoelectric coupling, and the used HPT method is an effective and promising method for obtaining new functional materials consisting of close-packed, ultra-dispersed, and exchange-interacting FE and FM phases. Moreover, the degree distribution of fractions, their size, and the strength of exchange interactions can be controlled by both high pressure and the turning number. The obtained new composites can be used for the creation of electronic devices for monitoring ultrahigh pressure and in integrated circuits of high-speed computing systems as a result of a decrease in the switching energy of nanodevices.

**References**

[1] a) W. Eerenstein, N. Mathur, J. F. Scott, *nature* **2006**, 442, 759; b) A. C. Garcia-Castro, Y. Ma, Z. Romestan, E. Bousquet, A. H. Romero, *Advanced Functional Materials* **2021**.

[2] J. Ma, J. Hu, Z. Li, C. W. Nan, *Advanced materials* **2011**, 23, 1062.

[3] N. Ortega, A. Kumar, J. Scott, R. S. Katiyar, *Journal of Physics: Condensed Matter* **2015**, 27, 504002.

[4] M. M. Vopson, *Critical Reviews in Solid State and Materials Sciences* **2015**, 40, 223.

[5] T. Rupp, B. D. Truong, S. Williams, S. Roundy, *Materials* **2019**, 12, 512.

[6] W. Huang, Y. Liu, Z. Luo, C. Hou, W. Zhao, Y. Yin, X. Li, *Journal of Physics D: Applied Physics* **2018**, 51, 234005.

[7] a) J. Zhai, Z. Xing, S. Dong, J. Li, D. Viehland, *Applied physics letters* **2006**, 88, 062510; b) G. Mioni, S. Grondin, L. Bardi, F. Stablum, *Behavioural brain research* **2020**, 377, 112232.

[8] V. Pashchenko, A. Khor'yakov, A. Pashchenko, Y. S. Prilipko, A. Shemyakov, *Inorganic Materials* **2014**, 50, 191.

[9] M. Bersweiler, P. Bender, L. G. Vivas, M. Albino, M. Petrecca, S. Mühlbauer, S. Erokhin, D. Berkov, C. Sangregorio, A. Michels, *Physical Review B* **2019**, 100, 144434.

[10] Y. O. Tykhonenko-Polishchuk, A. Tovstolytkin, *Journal of Nano-and Electronic Physics* **2017**, 9, 2028.

[11] M. Zhu, Z. Zhou, B. Peng, S. Zhao, Y. Zhang, G. Niu, W. Ren, Z. G. Ye, Y. Liu, M. Liu, *Advanced Functional Materials* **2017**, 27, 1605598.1.

[12] P. Palmero, *Nanomaterials* **2015**, 5, 656.

[13] S. B. Waje, M. Hashim, W. D. W. Yusoff, Z. Abbas, *Applied Surface Science* **2010**, 256, 3122.

[14] a) A. Pashchenko, **1986**; b) V. Pashchenko, A. Nesterov, Y. G. Drigibka, G. Potapov, E. Khapalyuk, A. Shemyakov, V. Berezhnaya, *Powder Metallurgy and Metal Ceramics* **1994**, 33, 300; c) V. Pashchenko, A. Nesterov, Y. G. Drigibka, G. Potapov, E. Khapalyuk, A. Shemyakov, V. Berezhnaya, *Powder Metallurgy and Metal Ceramics* **1995**, 33, 300.

[15] a) Y. Ivanisenko, R. Kulagin, V. Fedorov, A. Mazilkin, T. Scherer, B. Baretzky, H. Hahn, *Materials Science and Engineering: A* **2016**, 664, 247; b) R. Kulagin, Y. Beygelzimer, Y. Ivanisenko, A. Mazilkin, H. Hahn, *Procedia engineering* **2017**, 207, 1445.

[16] A. Pashchenko, V. Pashchenko, Y. F. Revenko, V. Spuskanyuk, N. Kasatka, V. Turchenko, A. Shemyakov, *Technical Physics Letters* **2010**, 36, 566.

[17] R. Blaak, *The Journal of Chemical Physics* **2000**, 112, 9041.

[18] H. M. Rietveld, *Journal of applied Crystallography* **1969**, 2, 65.

[19] J. Rodríguez-Carvajal, *Physica B: Condensed Matter* **1993**, 192, 55.

[20] Z. Wei, A. Pashchenko, N. Liedienov, I. Zatovsky, D. Butenko, Q. Li, I. Fesych, V. Turchenko, E. Zubov, P. Y. Polynchuk, *Physical Chemistry Chemical Physics* **2020**, 22, 11817.

[21]     a) A. Patterson, *Physical review* **1939**, 56, 978; b) B. Warren, *Journal of applied physics* **1941**, 12, 375.

[22]     A. Abramov, D. Alikin, V. Yuzhakov, A. Nikitin, S. Latushko, D. Karpinsky, V. Y. Shur, A. Kholkin, *Ferroelectrics* **2019**, 541, 1.

[23]     A. Okada, Y. Yamamoto, T. Yoshiie, I. Ishida, K. Hamada, E. Hirota, *Materials Transactions, JIM* **1993**, 34, 343.

[24]     Z. Min, Z. Zi, Q. Liu, Z. Peng, X. Tang, Y. Jie, X. Zhu, Y. Sun, J. Dai, **2013**.

[25]     M. A. N. Varshavskii M. T. PVP, Suntsov N.V., Miloslavskii A.G. , **1988**.

[26]     S. Mørup, M. F. Hansen, C. Frandsen, *Beilstein journal of nanotechnology* **2010**, 1, 182.

[27]     P. Suresh, S. Srinath, *Journal of alloys and compounds* **2013**, 554, 271.

[28]     D. Albrecht, S. Lisenkov, W. Ren, D. Rahmedov, I. A. Kornev, L. Bellaiche, *Physical review* **2010**, 81, 140401.1.

[29]     a) H. FD., **2009**; b) H. M. L., *MaDC-F* **2020**.

[30]     K. P. Belov, *Physics-Uspekhi* **1996**, 39, 623.

[31]     A. Guimaraes, Springer-Verlag Berlin Heidelberg: Wiley-IEEE Press, 2009.

[32]     S. Krupička, *Physik der Ferrite und der verwandten magnetischen Oxide*, Springer-Verlag, **2013**.

[33]     A. Perejon, P. E. Sanchez-Jimenez, L. A. Perez-Maqueda, J. M. Criado, J. R. de Paz, R. Saez-Puche, N. Maso, A. R. West, *Journal of Materials Chemistry C* **2014**, 2, 8398.

[34]     Y. M. Gufan, V. M. Kalita, *Fizika Tverdogo Tela* **1987**, 29, 3302.

[35]     G. Y. Lavanov, V. Kalita, V. Loktev, *Low Temperature Physics* **2014**, 40, 823.

[36]     G. R. Hoogeboom, T. Kuschel, G. E. Bauer, M. V. Mostovoy, A. V. Kimel, B. J. van Wees, *Physical Review B* **2021**, 103, 134406.

[37]     a) R. Levitin, A. S. Markosyan, *Soviet Physics Uspekhi* **1988**, 31, 730; b) X. X. Shi, X. Q. Liu, X. M. Chen, *Advanced Functional Materials* **2017**, 27, 1604037.

[38]     Z. Gareeva, A. Zvezdin, L. Kalyakin, T. Gareev, *Journal of Magnetism and Magnetic Materials* **2020**, 515, 167255.

[39]     H. Dixit, J. H. Lee, J. T. Krogel, S. Okamoto, V. R. Cooper, *Scientific reports* **2015**, 5, 1.

[40]     J.-G. Park, M. D. Le, J. Jeong, S. Lee, *Journal of Physics: Condensed Matter* **2014**, 26, 433202.

[41]     H. A. Haus, J. R. Melcher, *Electromagnetic fields and energy*, Prentice Hall Englewood Cliffs, NJ, **1989**.

[42]     A. Pashchenko, N. Liedienov, Q. Li, I. Makoed, D. Tatarchuk, Y. Didenko, A. Gudimenko, V. Kladko, L. Jiang, L. Li, *Materials Chemistry and Physics* **2021**, 258, 123925.

[43]     H. Naganuma, Y. Inoue, S. Okamura, *Journal of the Ceramic Society of Japan* **2010**, 118, 656.

[44]     a) A. Pashchenko, N. Liedienov, Q. Li, D. Tatarchuk, V. Turchenko, I. Makoed, V. Y. Sycheva, A. Voznyak, V. Kladko, A. Gudimenko, *Journal of Magnetism and Magnetic Materials* **2019**, 483, 100; b) N. Liedienov, A. Pashchenko, V. Turchenko, V. Y. Sycheva, A. Voznyak, V. Kladko, A. Gudimenko, D. Tatarchuk, Y. V. Didenko, I. Fesych, *Ceramics International* **2019**, 45, 14873; c) I. I. Makoed, N. A. Liedienov, A. V. Pashchenko, G. G. Levchenko, K. I. Yanushkevich, *Journal of Alloys and Compounds* **2020**, 155859.

[45]     E. Elayaperumal, G. Murugesan, M. Malathi, *Materials Letters* **2021**, 300, 130048.

[46]     R. Ramesh, L. Martin, *La Rivista del Nuovo Cimento* **2021**, 1.

[47]     V. V. Zhirnov, R. K. Cavin, *Nature Nanotechnology* **2008**, 3, 77.

[48]     a) J. Lv, X. Lou, J. Wu, *Journal of Materials Chemistry C* **2016**, 4, 6140; b) X. Ming, X. Meng, Q.-L. Xu, F. Du, Y.-J. Wei, G. Chen, *RSC advances* **2014**, 4, 64601.

[49] F. Kubel, H. Schmid, *Acta Crystallographica Section B: Structural Science* **1990**, 46, 698.

[50] a) G. Zhang, F. Liu, T. Gu, Y. Zhao, N. Li, W. Yang, S. Feng, *Advanced Electronic Materials* **2017**, 3, 1600498; b) C. Xu, Y. Li, B. Xu, J. í?iguez, W. Duan, L. Bellaiche, *Advanced Functional Materials* **2016**, 27, 1604513; c) J. Chen, B. Xu, X. Q. Liu, T. T. Gao, X. M. Chen, *Advanced Functional Materials* **2018**, 29.

[51] a) A. A. Belik, H. Yusa, N. Hirao, Y. Ohishi, E. Takayama-Muromachi, *Chemistry of Materials* **2009**, 21, 3400; b) V. Khomchenko, D. Karpinsky, A. Kholkin, N. Sobolev, G. Kakazei, J. Araujo, I. Troyanchuk, B. Costa, J. Paixao, *Journal of Applied Physics* **2010**, 108, 074109.

[52] D. C. Arnold, *IEEE transactions on ultrasonics, ferroelectrics, and frequency control* **2015**, 62, 62.

[53] D. Damjanovic, *Applied Physics Letters* **2010**, 97, 062906.

[54] C. S. Knee, M. G. Tucker, P. Manuel, S. Cai, J. Bielecki, L. Börjesson, S. G. Eriksson, *Chemistry of Materials* **2014**, 26, 1180.

**Novel Multiferroic Nanocomposite with High Pressure-Modulated Magnetoelectric Coupling**

Chunrui Song[a], Nikita Liedienov[a,b], Igor Fesych[c], Roman Kulagin[d], Yan Beygelzimer[b], Xin Zhang[a], Yonghao Han[a,*], Quanjun Li[a], Bingbing Liu[a], Aleksey Pashchenko[a,b,e,*], Georgiy Levchenko[a,b,*]

[a] *State Key Laboratory of Superhard Materials, International Center of Future Science, Jilin University, Changchun 130012, P. R. China*
[b] *Donetsk Institute for Physics and Engineering named after O.O. Galkin, NAS of Ukraine, Kyiv 03028, Ukraine*
[c] *Taras Shevchenko National University of Kyiv, Kyiv 01030, Ukraine*
[d] *Institute of Nanotechnology, Karlsruhe Institute of Technology, Karlsruhe 76021, Germany*
[e] *Institute of Magnetism, NAS of Ukraine and MES of Ukraine, Kyiv 03142, Ukraine*

These authors contributed equally: Chunrui Song, Nikita Liedienov

*Corresponding author
*E-mail address*:  hanyh@jlu.edu.cn (Yonghao Han)
   alpash@ukr.net   (Aleksey Pashchenko)
   g-levch@ukr.net (Georgiy Levchenko)

**SM 1**

**X-ray patterns of the initial BLFO and MZFO powders at room temperature refined by the Rietveld method**

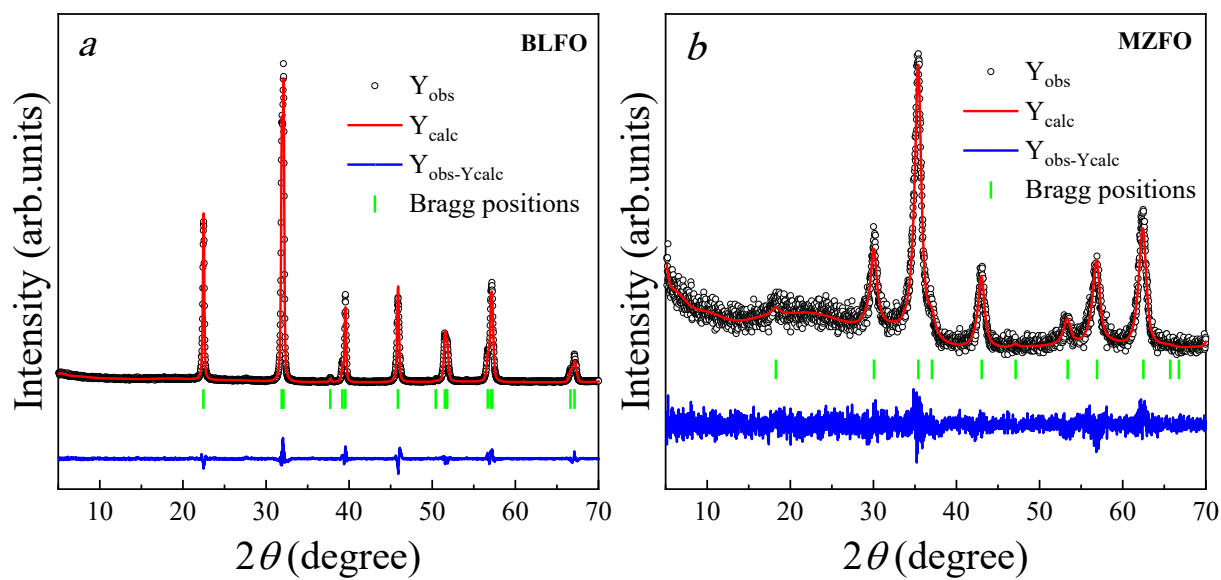

**Fig. SM1**. Rietveld refinement for the BLFO (*a*) and MZFO (*b*) powders fitted with observed (open circles) and calculated (line) XRD patterns, and the difference between the observed and calculated XRD intensities. The vertical bars indicate the angular positions of the allowed Bragg reflections.

Crystallographic parameters of the initial BLFO perovskite and MZFO spinel powders refined by Rietveld method.

| Sample | BLFO | MZFO |
|---|---|---|
| Crystal system | rhombohedral | cubic |
| Space group | $R3c$ (No. 161) | $Fd\bar{3}m$ (No. 227) |
| Lattice parameters | $a$ = 5.57321(40) Å $b$ = 5.57321(40) Å $c$ = 13.78585(100) Å $\alpha = \beta = 90°; \gamma = 120°$ | $a = b = c$ = 8.40195(229) Å; $\alpha = \beta = \gamma = 90°$ |
| Unit cell volume [Å$^3$] | 370.829(46) | 593.116(279) |
| Number of formula units $Z$ | 6 | 8 |
| X-Ray density [g/cm$^3$] | 8.217 | 5.238 |
| $R_p$ [%] | 7.91 | 11.50 |
| $R_{wp}$ [%] | 11.20 | 15.40 |
| $R_{exp}$ [%] | 7.52 | 14.22 |
| $\chi^2$ [%] | 2.24 | 1.17 |

**SM 2**

**Crystallite size determination**

The crystallite size of the composite was calculated by the X-ray line broadening method using the Scherrer formula [1]:

$$D_{hkl} = K \cdot \lambda / \beta_{hkl} \cdot \cos\theta_{hkl},$$

where $D_{hkl}$ in nm is the average crystallite size along the direction normal to the diffraction plane (hkl), $K$ is the shape factor equal to 0.9, $\lambda$ is the X-ray wavelength 0.15406 nm of Cu$_{K\alpha}$-radiation, $\beta_{hkl}$ is the integral breadth of the peak related to the diffraction plane (hkl), and $\theta_{hkl}$ is the Bragg angle in radians for the crystallographic plane (hkl). The true integral peak width was calculated using the Warren formula [2]:

$$\beta_{hkl} = \beta_{exp} - \beta_0,$$

where $\beta_{exp}$ is the experimental peak width of the sample at half maximum intensity; $\beta_0$ is the instrumental broadening of the diffraction line, which depends on the design features of the diffractometer. The Lorentz function was chosen to describe the shape of the diffraction peaks. In order to exclude the instrumental broadening $\beta_0$, a standard silicon Si X-ray powder diffraction data is recorded under the same condition and is eliminated from the observed peak width. The (012) peak of perovskite phase and (311) peak of spinel phase were chosen for calculations as the most suitable for crystallite size determination (see Fig. SM2).

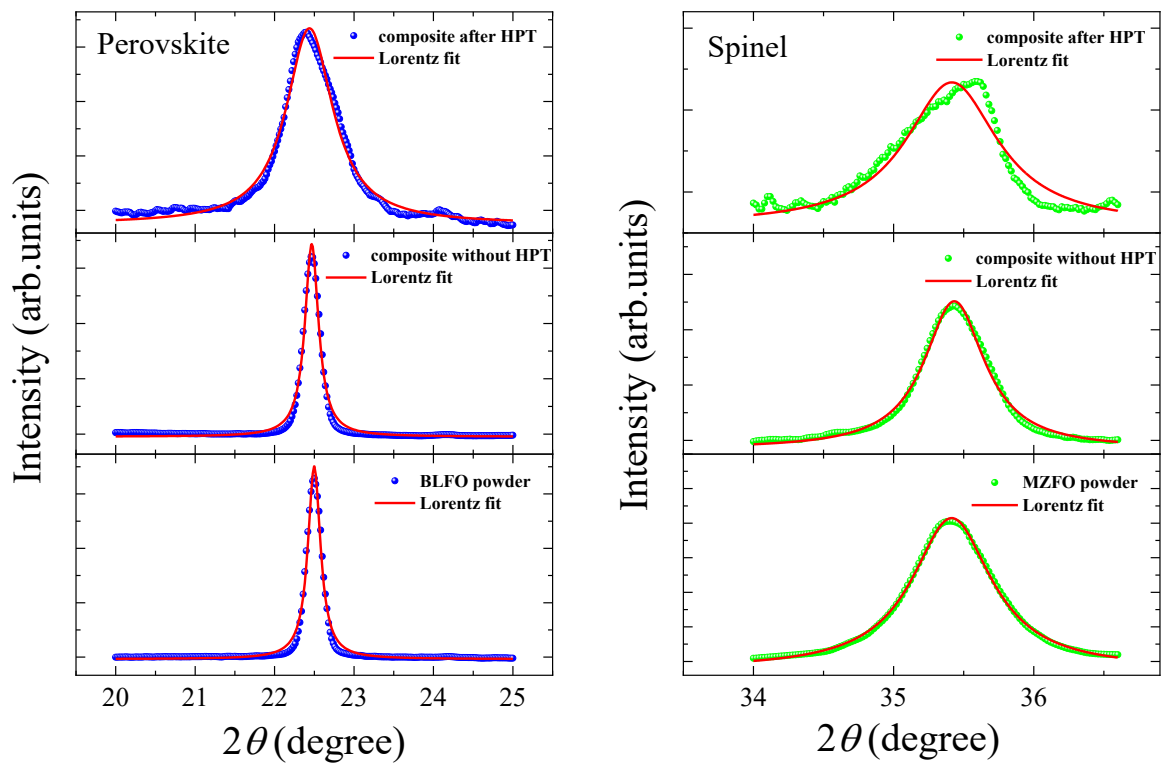

**Fig. SM2**. The (012) peak of perovskite phase (left) and (311) peak of spinel phase (right) fitted using the Lorentz function.

The average crystallite size of the perovskite phase was found to be approximately 131 ± 3.4 nm for the initial BLFO powder, 103 ± 2.8 nm for the BLFO–MZFO composite without HPT, and 14 ± 0.2 nm for the BLFO–MZFO composite after HPT (see Table SM2.1).

**Table SM2.1**

Experimental parameters of the Scherrer formula for determining average size of the coherent scattering regions $D_{012}$ in the perovskite phase.

| Sample | $2\theta$, degree | $\beta$, radian | $\cos\theta$ | $\lambda$, nm | K | $D_{012}$, nm |
|---|---|---|---|---|---|---|
| BLFO powder | 22.499 | 0.0011 | 0.981 | 0.15406 | 0.9 | 131 |
| composite without HPT | 22.468 | 0.0014 | 0.981 | 0.15406 | 0.9 | 103 |
| composite after HPT | 22.440 | 0.0102 | 0.981 | 0.15406 | 0.9 | 14 |

The average crystallite size of the spinel phase was found to be approximately 15 ± 0.1 nm for the initial MZFO powder, 15 ± 0.2 nm for the BLFO–MZFO composite without HPT, and 12 ± 0.5 nm for the BLFO–MZFO composite after HPT (see Table SM2.2).

**Table SM2.2**

Experimental parameters of the Scherrer formula for determining average size of the coherent scattering regions $D_{311}$ in the spinel phase.

| Sample | $2\theta$, degree | $\beta$, radian | $\cos\theta$ | $\lambda$, nm | K | $D_{311}$, nm |
|---|---|---|---|---|---|---|
| MZFO powder | 35.4142 | 0.0098 | 0.953 | 0.15406 | 0.9 | 15 |
| composite without HPT | 35.4330 | 0.0095 | 0.953 | 0.15406 | 0.9 | 15 |
| composite after HPT | 35.4148 | 0.0117 | 0.953 | 0.15406 | 0.9 | 12 |

**SM3**

### Particle size distribution

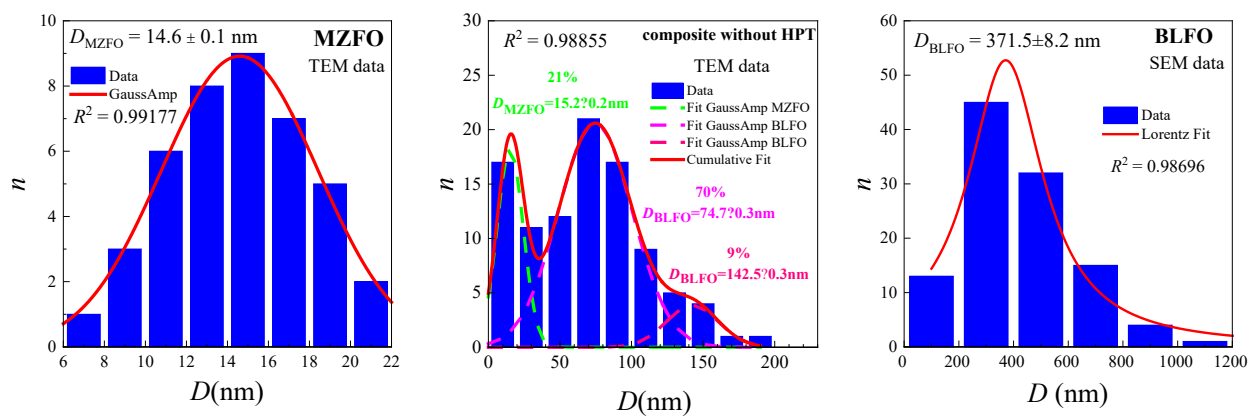

**Fig. SM3**. The average particle size $D$ and the size distribution for the initial MZFO (left, TEM) and BLFO (right, SEM) powders, as well as for the BLFO–MZFO composite without HPT (center, TEM). For the composite, the average particle sizes in the BLFO and MZFO fractions are indicated in their percentage.

The choice of the function for approximating the $n(D)$ experimental data was carried out using three fitting functions:

$$f(x) = A \cdot \exp\left[-\frac{(x-x_0)^2}{2\sigma^2}\right], \text{ Gaussian (GaussAmp)},$$

$$f(x) = \frac{2A}{\pi} \cdot \frac{\sigma}{4 \cdot (x-x_0)^2 + \sigma^2}, \text{ Lorentzian (Lorentz)},$$

$$f(x) = \frac{A}{\sqrt{2\pi}\sigma x} \cdot \exp\frac{-\left[\ln\frac{x}{x_0}\right]^2}{2\sigma^2}, \text{ Log-Normal Distribution},$$

here $A$ is a constant; $x_0$ is the expected value or average particle size, i.e. $D = x_0$; $\sigma^2$ is the dispersion of particles in size. The criterion for choosing the fitting function was the approximation accuracy when the determination coefficient $R^2$ takes the maximum value.

The percentage of the BLFO (79%) and MZFO (21%) powders in the BLFO–MZFO composite without HPT was obtained from the normalization condition of $\int_0^{+\infty} n(D)dD = 1$ for the fitting functions.

**SM4**

**Chemical composition based on the EDS study**

**Table SM4**

Chemical composition (at. %) according to EDS data in the initial BLFO and MZFO powders, as well as the BLFO–MZFO composites without and after HPT.

| Sample | Case | Element (at. %) | | | | | |
|---|---|---|---|---|---|---|---|
| | | Bi | La | Mn | Zn | Fe | O |
| BLFO powder | Stoichiometry | 18.00 | 2.00 | – | – | 20.00 | 60.00 |
| | EDS data | 16.62 | 2.05 | – | – | 27.92 | 53.41 |
| MZFO powder | Stoichiometry | – | – | 9.30 | 3.72 | 29.88 | 57.10 |
| | EDS data | – | – | 9.57 | 4.65 | 28.26 | 57.52 |
| Composite without HPT | Stoichiometry | 14.40 | 1.60 | 1.86 | 0.74 | 21.98 | 59.42 |
| | EDS data | 10.47 | 1.17 | 2.31 | 0.92 | 20.91 | 64.22 |
| Composite after HPT | Stoichiometry | 14.40 | 1.60 | 1.86 | 0.74 | 21.98 | 59.42 |
| | EDS data | 8.90 | 1.13 | 2.50 | 0.88 | 23.93 | 62.66 |

## Magnetic properties of the initial BLFO and MZFO powders

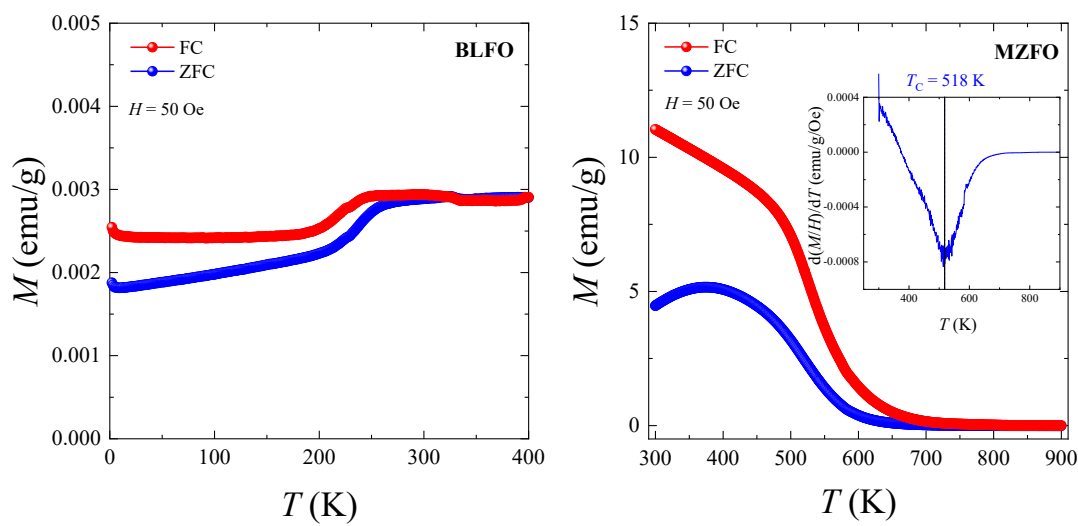

**Fig. SM5.1**. The $M_{ZFC}(T)$ and $M_{FC}(T)$ temperature dependencies for the initial BLFO and MZFO powders ($T_C$ is the Curie temperature, $M/H$ is the magnetic susceptibility).

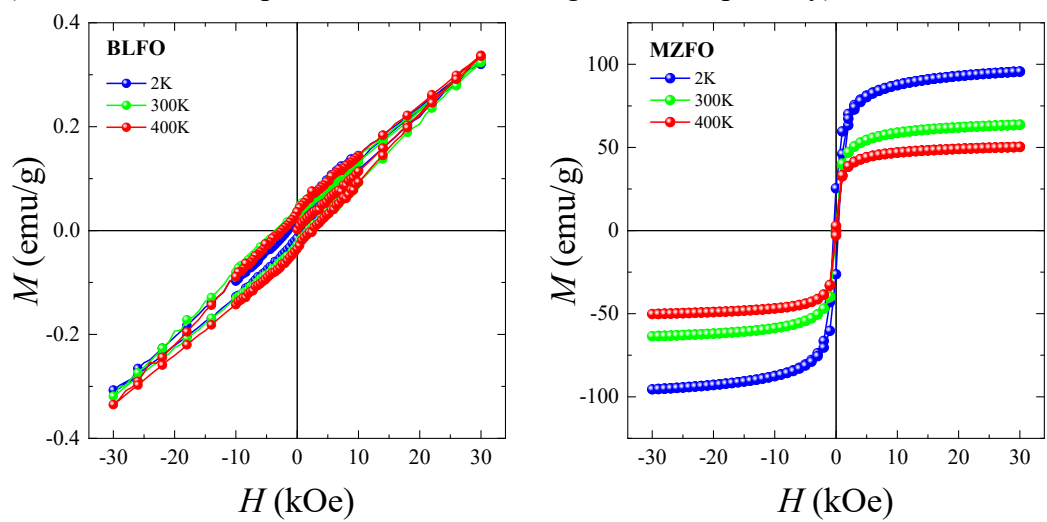

**Fig. SM5.2**. The field dependences of the magnetization $M(H)$ for the initial BLFO and MZFO powders.

# Calculation of ferromagnetic exchange length for the $Mn_{0.6}Zn_{0.3}Fe_{2.1}O_4$

The exchange length $l_{ex}$ in ferromagnets has the form [3]:

$$l_{ex} = \sqrt{\frac{2 \cdot A}{4 \cdot \pi \cdot M_S^2}},$$

where $A = \dfrac{J_{ex} \cdot S^2}{a}$ is the exchange stiffness constant in erg/cm;

$M_S$ is the saturation magnetization in emu/cm$^3$;

$a$ is the lattice constant in cm;

$J_{ex} = \dfrac{3 \cdot k_B \cdot T_C}{Z \cdot S(S+1)}$ is the exchange integral in erg;

$k_B = 1.38 \cdot 10^{-16}$ erg/K is the Boltzmann constant;

$S$ is the spin quantum number in $\hbar$;

$T_C$ is the Curie temperature in K;

$Z = 6$ is the number of nearest neighbors.

Considering the valence and filling coefficients of the *A*-tetrahedral and *B*-octahedral positions, the molar formula of the MZFO spinel is $\{Mn^{2+}_{0.6}Zn^{2+}_{0.3}Fe^{3+}_{0.1}\}_A \{Fe^{3+}_{1.9}Fe^{2+}_{0.1}\}_B O^{2-}_4$ [4].

$S(Mn^{2+}) = 5/2\ \hbar$, $S(Fe^{3+}) = 5/2\ \hbar$, $S(Fe^{2+}) = 4/2\ \hbar$

$S_A = (0.6 \cdot 5/2 + 0.1 \cdot 5/2)\ \hbar = 1.75\ \hbar$

$S_B = (1.9 \cdot 5/2 + 0.1 \cdot 4/2)\ \hbar = 4.95\ \hbar$

$S = (4.95 - 1.75)\ \hbar = 3.2\ \hbar$

Using experimental $T_C$ = 518 K, $M_S$ = 461 emu/cm$^3$, and $a = 8.40195 \cdot 10^{-8}/\sqrt{2} = 5.941 \cdot 10^{-8}$ cm for the Mn$_{0.6}$Zn$_{0.3}$Fe$_{2.1}$O$_4$ spinel, the exchange integral $J_{ex}$, the exchange stiffness constant $A$, and the exchange length $l_{ex}$ have been obtained:

$J_{ex}$ = 2.659·10$^{-15}$ erg,

$A$ = 4.584·10$^{-7}$ erg/cm,

$l_{ex}$ = 5.94 nm.

**References**

[1] Klug, H. P. Alexander, E. Leroy, *X-Ray Diffraction Procedure*, X-Ray Diffraction Procedure, **1954**.
[2] A. L. Patterson, *Physical Review* **1939**, 56, 978.
[3] a) A. P. Guimares, *Principles of Nanomagnetism*, Principles of Nanomagnetism, **2009**; b) Y. O. Tykhonenko-Polishchuk, A. I. Tovstolytkin, *Journal of Nano- and Electronic Physics* **2017**, 9, 02028.
[4] V. P. Pashchenko, A. A. Khor'Yakov, A. V. Pashchenko, Y. S. Prilipko, A. A. Shemyakov, *Inorganic Materials* **2014**, 50, 191.